Original Article

# Interactions between eagles and semi-domestic reindeer – lessons learned from field surveys and deterrents

Aemilius Johannes van der Meiden, Andrés López-Peinado, Peter Sunesson, Christian Emilsson, Navinder J Singh*

*Department of Wildlife, Fish and Environmental Studies, Faculty of Forest Sciences, Swedish University of Agricultural Sciences, Umeå, SE 90183*

*Corresponding author: navinder.singh@slu.se

**Abstract**

Predation by eagles on semi-domesticated reindeer (*Rangifer tarandus*) is an emerging human wildlife conflict in Fennoscandia. Both the Golden (*Aquila chrysaetos*) and the White-tailed eagle (*Haliaeetus albicilla*) are believed by herders to predate on reindeer, however, there is a considerable knowledge gap regarding extent of predation and scavenging by each species, and their distribution and behaviour within the reindeer herding areas. Lethal and non-lethal methods have been suggested to reduce this conflict with eagles. We developed this project to fill the existing knowledge gaps by investigating the patterns of eagle abundance before, during, and after reindeer calving in a herding district in northern Sweden, and testing the effect of two potential deterrents (air ventilators and rotating prisms) in diverting eagles away from reindeer calving areas. During the single study period, we made 12, 47, and 17 eagle observations before, during, and after calving respectively. Out of these observations, 34 were of Golden eagles, 33 of White-tailed eagles, and for 9 observations the species could not be confirmed. Eagle abundance increased during calving and decreased again after calving ended. No attacks by eagles on calves were observed. The odds of observing eagles were significantly higher in the control area compared to areas with deterrents. More sub-adults were observed during calving, and both species were present in the area. The extent of predation was difficult to infer using direct observations and deterrents seem to show promise in diverting eagles away from calving grounds. These studies should be replicated to get a general picture of the issue and testing the efficiency of deterrents in diverting eagles away from reindeer across reindeer herding districts.

# Introduction

With predators declining globally over the past two centuries, the conservation of these species has become more important than ever (Ripple et. al, 2014). Conservation of predators however often clashes with human livelihoods, land use, and behaviour, for instance in the form of fear, lethal encounters, or predation on livestock or pets (Vittersø et. al, 1998; Linnell et al., 2002; Olivera-Méndez et. al, 2019). One of these issues is predation on semi domestic reindeer (*Rangifer tarandus*) by large predators in Fennoscandia. Predators such as Bear (*Ursos arctos*), Lynx (*Lynx lynx*), Wolverine (*Gulo gulo*), Wolf (*Canis lupus*), and Golden eagle (*Aquila chrysaetos*) are known to prey on reindeer calves, juveniles, and sometimes even adults (Nybakk, 1999; Norberg et al., 2006; Horstkotte et. al, 2022). Furthermore, there are rising concerns among reindeer herders that White-tailed eagles (*Haliaeetus albicilla*) are also a part of the issue (Ekblad et al, 2020). The true extent of eagle predation, especially determining whether it killed an individual, or merely scavenged, are difficult to establish since observations are lacking and systematic data is difficult to obtain (Hjernquist, 2011; Mattisson, 2018).

The current population of Golden eagles in Sweden has fluctuated around 1500 individuals between 2006 and 2021, but the mortality of eagles from traffic accidents is also at an all-time high (Åsbrink & Hellström, 2022, Singh et al. 2024 ). Therefore, simply relying on breeding parameters does not reflect the true picture of the eagle population. The two northernmost counties hold most of the eagle territories, with more scattered populations in the south, and a noticeable population on the island of Gotland (Moss 2015 , Åsbrink & Källman, 2023). Predation on semi-domestic reindeer by Golden eagles has been documented, but the balance between direct attacks on reindeer and scavenging on animals that are already dead is unknown. The estimated percentage of annual predation by eagles ranges from 0 - 4.2% (Nybakk et. all, 1999; Norberg et. al, 2006; Nieminen et. al, 2010). Nybakk et. al (1999) showed a predation rate on calves to be 2.4% in August, when the calves are already around two to three months old. Predation on neonates is claimed by herders to be the most serious problem, since that is when calves are most vulnerable (Tjernberg, 1981). This has recently been shown for brown bears, and most of the predation occurred between the $1^{st}$ of May and $9^{th}$ of June (Støen et. al, 2022).

The White-tailed eagle has recently recovered in the region after an unprecedented low from the negative effects of DDT in the 1900's (Helander et. al. 2008), and according to herders, their presence is increasingly observed in the reindeer herding areas attacking calves and scavenging on carrion. This is leading to rising demands among the reindeer herders for their inclusion in the annual compensation for predation (Ekblad, 2020). Sightings from Scotland report that White-tailed eagles may kill red deer calves (Love, 2013), and in Norway they are known to predate on sheep. Regarding reindeer, White-tailed eagles are known to scavenge on dead individuals, both calves and adults, but attacks have not been reported (Mattisson, 2018). Although the literature suggests that White-tailed eagles are of no concern to reindeer herders, they may possibly influence it indirectly by reducing the amount of carrion available for the Golden eagles, forcing them to hunt more frequently (potentially on reindeer). More knowledge on eagle species status, distribution and behaviour is therefore required to form a basis for the predation compensation scheme and further management decisions relating to predator management (Mattisson et. al, 2018).

Reindeer females usually aggregate into large herds and give birth from mid-May to early June (Singh et al. 2021). During this period both cows and calves are vulnerable to predation, with newborns being easy targets for predators and females building up their lost body reserves during winter and giving birth (Linnell, 1995). The Golden eagle, especially juveniles and subadults, may migrate hundreds of kilometers to be spatially and temporally coupled with calving reindeer (Singh et al., 2021). The proportion of reindeer in their diet can vary from 6% to 43% (Mattisson, 2018), and they may scavenge on afterbirths, stillborn calves, or dead reindeer but have also been reported to attack young reindeer (Nybakk et al., 1999). True numbers of eagle predation, especially determining whether it killed an individual, or merely scavenged are difficult to quantify (Hjernquist, 2011; Mattisson, 2018). Predation on neonates is claimed by herders to be the most serious problem, since that's when calves are most vulnerable (Tjernberg, 1981; Nybakk, 1999). The compensation system for predation on Reindeer by eagles involves a payment of 1 million SEK annually to whole the Sami community comprising of 51 herding villages or varying size, with the amount then split between the herding districts proportional to the size of the herds in their respective villages (Sametinget, 2023). The White-tailed Sea eagles are currently not regarded in this scheme.

Deterrents are often used as a non-lethal means to mitigate human-wildlife conflicts, using the concept of ecology of fear, aiming to modify animal behavior to minimize interactions or prevent damage (Chapron et al., 2014; Miller et al., 2016). Various types of deterrents have been employed across different species and contexts. These can range from simple alarms to more sophisticated devices emitting predator calls or distress signals (Harris & Davis, 1998). There may be devices or materials that visually disrupt animals, such as scarecrows, flashing lights, or reflective tapes. These exploit animals' natural instincts to avoid potential threats (Harris & Davis, 1998). Chemical substances that emit odors or tastes unpleasant to animals, including capsaicin-based sprays for mammals or bitter-tasting coatings for crops (Baylis et al., 2012, Cummings et al., 1992). Fences, netting, or other structures that physically prevent animals from accessing certain areas or resources or repellents comprising substances applied to surfaces or crops that repel animals through smell or taste, without causing harm. Examples include predator urine or garlic-based sprays (Nolte & Wagner, 2000). Electric fences or mats that deliver non-lethal shocks to deter animals from crossing boundaries (McDonald et al., 1981), and finally, introducing natural predators or competitors to discourage certain species from frequenting human settlements or agricultural areas (Yadav & Kumar, 2021).

The effectiveness of deterrents varies depending on factors such as the target species, local environment, and the specific situation (Bishop et. al, 2003; Levin et. al, 2008). For birds, especially raptors, which make great use of their excellent eyesight for looking for food and navigate themselves (Mitkus et al., 2018). Reducing or disturbing their sight has a potential as a deterrent method, as shown by reflective devices used to avoid bird collisions with cables and fences (Barrientos et al., 2011). Nonetheless, not a lot of research has been done on raptors regarding mirror/reflective devices, but there have been some successful trials (Bishop et. al, 2003; Levin et. al, 2008). It's essential to consider the ethical and ecological implications of deterrent use, ensuring that they do not cause harm to animals or disrupt natural behaviors. Additionally, long-term effectiveness may require periodic changes in deterrent strategies to prevent habituation by wildlife.

We developed this study to fill the above knowledge gaps on eagle predation on reindeer issue by producing objective observational data and testing a pilot project involving the use of deterrents to divert eagles away from reindeer during the calving season. The specific objectives were to investigate the patterns of abundance of both Golden and White-tailed eagles before, during, and after reindeer calving in a reindeer herding district in northern Sweden. We tested the effect of two potential deterrents (air ventilators and rotating prisms) in diverting eagles away from reindeer calving areas. These non-lethal devices rely on reflecting sunlight in an upward angle to deter birds from certain areas. Based on herders' reports, we expected both the Golden eagle and the White-tailed eagle to be present in the area. We expect the number of eagle observations of both species to increase during calving when compared to before, and after calving as per Singh et al. 2021, who showed that eagles may migrate to reindeer calving areas in spring. We expect that during calving the number of observations of subadults will increase and decrease after calving as it is the non-breeding individuals that primarily migrate (Ecke et al. 2017, Singh et al., 2021). Finally, if the deterrents work, we expect the amount of eagle observations to be lower in treatment areas compared to controls with no deterrents.

## Study area and Methods

The study area is located in the Vilhelmina Norra sameby (65.37°N, 14.64°E). Vilhelmina Norra is a reindeer herding area in northwestern Sweden that borders Norway (Figure 1). The total herding district is over 14.000 km$^2$, whereas our study area is the intensive calving area of size about 250 km$^2$ where the reindeer females aggregate to give birth in spring. The population target for this district is 11.000 reindeer for the whole area (Sametinget, 2023). The study area is characterized by rolling mountains of approximately 800 to 1300 m.a.s.l. During most months of the year, the area is covered in a thick pack of snow and the temperature during the study period ranged between -20C to 6C and wind speeds between 0-15m/s. Vegetation consists predominantly of birch (*Betula pubescens*) forests, Norway spruce (*Picea abies*), and Scots pine (*Pinus sylvestris*) trees, ground vegetation consists of subshrubs like crowberry (*Empetrum nigrum*) and blueberry (*Vaccinium myrtillus*).

*Study species*

The Golden eagle is one of Scandinavia's largest birds of prey, only surpassed by the White-tailed eagle in size (Cramp, 1992). The species occurs in most parts of Scandinavia, and there is great overlap with the areas where reindeer herding is prominent (Nieminen, 2011). Golden eagles face serious threats like lead poisoning (Legagneux et. al, 2014; Ecke et. al, 2017; Helander et. al, 2021), illegal shooting (Palmer, 1988), habitat loss, and human disturbances (Watson, 1997). Besides small mammals, birds, fish, reptiles, their diet partly consists of reindeer (Cramp, 1992; Tjernberg, 1981). The hunting technique used by Golden eagles to catch large prey, like the reindeer, especially in snowy conditions, is described by Watson (1997) as the 'glide attack with tail chase'. The White-tailed eagle is the largest raptor in Scandinavia (Cramp, 1992). It occurs throughout the whole of Sweden and Northern Europe (Artfakta, 2022). Persecution of the raptors, urbanization, toxins, and the forestry industry were the main reasons for their population decline (Helander, 2008). The White-tailed eagle, especially subadults, are more sociable compared to Golden eagles. Aggregation of immature individuals around food sources is not uncommon, and although Golden eagle are dominant, these aggregations can quickly reduce the food availability of an area (Mattisson, 2018). More than 90% of White-tailed eagle's diet is based on fish and birds (Ekblad et al, 2020), and the little percentage that mammals represent is usually related to scavenge due to the arrangement of claws on White-tailed eagle's feet, as it seems to make grappling ineffective when trying to grab a big mammal (e.g., reindeer) since they are not designed for catching larger prey (Mattisson, 2018).

*Data collection*

The field work was performed before calving (25/4/2022 – 29/4-2022), during calving (20/5/2022 – 26/4/2022), and after calving (26/6/2022 – 1/7/2022). Traditionally most calving happens during the second week of May (14/5 - 20/5) (Paoli, 2018). The before-during-after set-up allowed us to set a baseline of the number of eagles in the area before migrating individuals arrived. The after period allowed us to record if and how long a higher number of eagles remained in the area, to better assess the extent of the issue. Observations were done from 08:00 hours until 17:00 hours, as this is estimated to be the most active period for eagles. We used 8 · 65x Spotting scopes and 8 · 10x binoculars observations. The observations were conducted from vantage points selected for the highest visibility and area coverage. Vantage points were identified based on a viewshed analysis performed in QGIS with the Viewshed plug-in. To reach these places a snowmobile or skiis were used, but some places were only accessible by foot due to the terrain. Observers rotated between points throughout the weeks daily to randomize observer bias. Point count methods as well as territory mapping (in the first period) were used to determine the number of eagles in the area. When an eagle was spotted, its species, estimated age, flight path, behavior, observer, and time of observation were recorded. Flight paths were drawn on maps while in the field on a 1:50000 scale map. In case of uncertainty in age (plumages have transition states that are hard to distinguish) a range was given, 3-4 years for instance, or a plain 'adult' or 'subadult'. The accurate age of the individuals was in some cases not possible to establish due to the distance between the observers and the eagles (sometimes >5km), light conditions, or plumage transition states. Behaviour was categorized into 'sitting', 'flying' and 'hunting'. 'Hunting' was identified when an eagle circled above a herd and dived down onto an individual or a group of reindeer.

*Deterrents*

We tested the effect of two potential repelling devices: wind turbine ventilators and the Peaceful Pyramid (London, UK) later referred to as 'prisms' (Figure 1). The rotating, triangular, prisms are 12 cm high, run through a 12 V motor, and are powered by a 12 V car battery. The angle of the mirrors throws reflections obliquely upwards and thus scares the birds that fly into a field. This device was recommended by the Swedish Wildlife Damage Centre. The wind ventilators are the typical ventilators seen on rooftops. Sunlight is constantly reflected on the rotating metal ventilator, possibly repelling eagles like the prisms, just bigger and sturdier and driven by wind instead of a battery. The study area was split into three areas, one allocated to ventilators, one to prisms, and one control area (Figure 1). Twenty ventilators and seventeen prisms were placed in the study area separate from each other (Figure 1). The devices have been carefully placed in places with the highest visibility, predominantly on mountain tops. There was at least 1km between each deterrent. The devices were placed just before the 'during' calving period. No devices were present before calving, nor after. The effect of the deterrents was only tested during calving period to prevent habituation by eagles.

Admittedly, the area was divided in a way, where keeping different treatments independent of each other was not entirely possible. Hence, eagles had the possibility to cross over and range over all three areas. Nevertheless, due to the limitation that all reindeer of this village were restricted mainly to this area, there was no possibility to have control areas away from the treatment areas. We also assumed that the reindeer were randomly distributed across the entire study area at the time of study, and despite constant movements and fission – fusion of herds, the density was homogeneous across the study area during the periods of study.

Analysis was performed in R (R Core Team, 2021). To analyze the effect of the repelling devices, Generalized Linear Models with a Poisson distribution were built through the R package lme4 (Bates et al., 2015), and compared to a null model with an Omnibus test. Amount of observations was used as our response variable. "Day" was tested for effect and showed no effect on the number of eagles. "Day" was then included as a random effect in the model to account for daily fluctuations in for instance weather. This analysis was performed for control, ventilator, and prisms, but also for control vs. treatment.

All work was conducted under Ethical Permits No A11-2019 from the Swedish Agricultural Board - Jordbruksverket and Research Permit No. NV-07710-19 of the Swedish Environmental Protection Agency - Naturvårdsverket.

# Results

A total of 75 eagle observations were made during the study (Figure 2). Of these 75 observations, 34 were of Golden eagles, 32 of White-tailed eagles, and for 9 observations the species could not be identified. The number of observations was highest during the calving period for both eagle species (Figure 3). The number of observations was lowest in the 'before' calving week. The 'after' calving week had slightly more observations than before calving, but fewer than during calving. According to species, the number of White-tailed eagles' observations experienced a noticeable increase during calving, more pronounced than the change detected in the number of Golden eagle observations (Figure 3).

*Age distribution*

The number of adult Golden eagle observations remained stable across the study period (Figure 3). However, during calving, the number of subadult Golden eagle observations increased and declined again in the 'after' calving period. The White-tailed eagle observations reached their peak during calving, with both subadult and adult observations increasing and remaining higher in the after-calving period.

*Behaviour*

Expectedly, eagles were mostly observed while flying since that's when they are easiest to detect. On other occasions, they were also found sitting, with two observed cases of hunting (Figure 3).

*Treatment*

The likelihood of observing an eagle was significantly higher in the control area when compared to the deterrent areas (ventilator and prism) (Table 1, Figure 4). The likelihood was 1.7 and 1.1 times lower for ventilator and prism respectively (Table 1). The model based on treatment explained more than 41% of the variation in the eagle observations in the study area based on $R^2$. When taken together, the control area showed a 2.2 times higher odds of eagle occurrence compared to the treatment areas.

# Discussion

We produced several important results in the context of eagle reindeer interactions which provide crucial baseline information for management. The eagle abundance increased during the calving period and declined afterwards, more subadults were observed during calving, and both eagle species were present in the area. The extent of predation was difficult to infer using direct observations and deterrents seem to show promise in diverting eagles away from calving grounds. Experiments are a valuable tool in ecology to understand cause-and-effect relationships (Tilman, 1989). Nonetheless, experiments on wild animals in their natural habitat can be expensive, difficult to control and effort intensive, and therefore not many of these experiments have been performed in situations like ours (Fuller & Mosher, 1981; Hull et al. 2010). This study allowed us to delve deeper into the eagles' behaviour, both inter- and intraspecific, response to herders, reindeer, and the effect of the two types of repelling devices. With the before-during-after experimental set-up, patterns could be recognized in terms of species abundance, age class distribution (Stewart-Oaten et al., 1986).

    For Golden eagles, the monitoring of known territories was done in addition to direct observations from vantage points, to gather more information about the number of resident eagles in the area (Gregory, 2004; Singh, unpublished data). Nevertheless, the residency status of eagles was in most cases impossible to determine due to for instance variation in plumage per year, or distance from observer (Madge, 1979; Bibby, 2000).

    During the study, the number of observations fluctuated between the three periods. There was an increase in observations during calving, especially for the White-tailed eagles.

Their observations increased for both adults and subadults during calving. The increase in these observations could mean that food availability from reindeer is possibly more important to White-tailed eagles than to the Golden eagles. Nesting White-tailed eagles do not actively seek out reindeer carcasses (Ekblad et al, 2020), but will exploit them when encountered. For non-breeders, this information is unknown, but it seems plausible that non-breeders that are migrating into the area (from either Norwegian or Swedish side) to opportunistically seek out reindeer in a similar way as the Golden eagle (Singh et al. 2021). There are no known territories of White-tailed eagles in the area, suggesting the birds coming in the area are transient eagles, either subadult or non-breeding adults. All age classes were represented in the study period. As expected, most eagle observations were made during the calving period. Before calving, predominantly the Golden eagles were observed and the observations were relatively stable during the whole study period. If adult eagles had entered the area they would most likely be non-breeding individuals, since the territories are already occupied (Verhulst & Nilsson, 2008). Subadult Golden eagles are nomadic and are constantly looking for feeding opportunities, possibly explaining their slight increase during calving.

Two types of deterrent devices were tested during this study. The prisms have previously been successfully used to exclude birds from certain areas (Levin, 2008). The control area showed a significantly higher eagle observation rate when compared to treatment areas, even though the eagles had easy access to these areas. The deterrents seem to be a promising mechanism to divert eagles from certain areas. Although there were no significant differences between the two deterrent methods, our results seem to support that ventilators could be preferred due to no need for external power. The self-propelled shiny wind ventilators could make for a more sustainable alternative for the prisms, provided the birds are not allowed to be habituated and the spillover effects from eagles moving to other reindeer herding villages are controlled. The ventilators are sturdier, and since wind is an almost constant factor in the Swedish mountains, they are self-sufficient, replacing the need for batteries.

No direct predation from eagles was observed during this study. One dive was seen from a sub adult Golden eagle on a group of reindeer, possibly in an attempt to kill a laying calf, but to no success as the calf got up later and walked with the mother. White-tailed eagles were not observed hunting reindeer calves in our study, as in others (Mattisson, 2018; Ekblad, 2020). Predation rates can differ throughout years depending on fluctuations of availability of other prey species and scavenging opportunities (Sulkava, 1999). Besides reindeer, there are other alternative prey species available to hunt in the area such as mountain hare (*Lepus timidus*), fox (*Vulpes vulpes*), and willow (*Lagopus lagopus*) and rock ptarmigan (*Lagopus muta*). The abundance of after birth remains is another large food subsidy during this period that is undeniable for its importance to all scavenger species. Even long-tailed skuas (*Stercorarius longicaudus*) were observed feeding on these remains. Other studies like Nybakk et al, 1999, and Norberg, 2006 have found 12 cases of Golden eagle predation in over 850 mortality-sensing radio-collared reindeer, and 17 cases out of 621 radio-collared calves respectively, showing that predation can be challenging to record.

Causality and correlation are often mixed up in eagle predation and reindeer interactions. When an eagle is present on a reindeer calf carcass, it may seem obvious that it was killed by the particular eagle species. Predation is however only one of the options, and since eagles are likely to be one of the first predators to arrive at the site because of their soaring abilities (Peterson, 2001). It's likely eagles are already present before an observer or other predator could have spotted the calf, even though they had not killed the calf themselves. Predator misidentification is another issue, and observer bias could lead to higher predation assumptions than are realistic (Duriez, 2019). On top of that, often the two eagle species are confused, possibly confusing the White-tailed eagles for the predation on reindeer by Golden eagles, or wrongly estimating abundances of a certain species (Bibby, 2000; Mattisson, 2018).

The usage of snowmobiles was kept to a minimum to ensure eagles, and/or reindeer would not behave differently due to our presence (Skarin et al., 2004; Steenhof et al., 2014). Since the peak of eagle activity (09:00- 15:00) fell inside observation hours, it was safe to assume that most attacks would have been recorded, had they occurred in front of the observers. That does not exclude the attacks that may have happened elsewhere which the observers may have missed as the area was very remote.

Golden eagles can migrate into calving areas from all across Sweden and Fennoscandia as was shown by the GPS data (Singh et al. 2021). The eagles migrate into the calving areas from far away and leave after calving in pursuit of other food sources (Singh et al, 2021). This shows that it is challenging to predict where eagles may come from and predict where they would migrate to. Complementary to the deterrent devices, other methods could also be used to reduce predation. Increasing herder presence around the calving females, to focus on safeguarding the reindeer during the daylight hours between 06:00- 18:00 could be extremely rewarding as it requires no external measures (Bell & Austin, 1985; Vickery & Summers, 1992). Human presence can act as the most effective deterrent for most animal species (Widen et al. 2022), however this requires effort by the herders involving several people simultaneously, to continuously watch the herds and follow the calving female groups. This would result in increased labour costs which may be difficult to achieve, as emigration from reindeer herding as a livelihood and lack of sufficient manpower is already a big challenge for the herding community.

Stein et al. (2022) in a study from Norway adopted GPS marked birds, both territorial and non territorial individuals, and followed their observed clusters of locations to detect predation events and diets and validated them with field visits. Their method seems to be effective for large prey detection (e.g. reindeer and sheep) with caveats linked to underestimation of prey species that are swallowed whole. Such methods hold promise provided there is financing available to follow enough numbers of individuals for different age classes and species. Eventually, predictive models from such datasets could allow a better prognosis of predation levels and hence estimations of compensation. Supplementary feeding has also been used as a conservation technique for the Spanish imperial eagle (Gonzalez, 2006), and hen harrier (Redpath, 2001). Diverting the eagles from calving areas through supplementary feeding could aid in reducing predation on reindeer (Knight & Anderson, 1990; McCollough et. al, 1994), but could also have adverse effects like disease transmission (Sorensen et al., 2014), attracting other predators (Pearson & Husby, 2021), and possibly turning into an ecological trap (Robb et al., 2008). With both eagle species migrating from all over Scandinavia, assessing the amount of food that needs to be put out, and the locations for feeding stations for this measure to be effective may be unrealistic. Moreover, there may be a risk of spillover effect on adjacent herding villages.

Predation on calves by Golden eagles early in the calving period is, according to the reindeer herders, the biggest problem. Therefore, devices must be set up just before the calving peak. To prevent habituation, these devices should only be used during calving. Most calving should have happened in two weeks, and habituation could be kept to a minimum. Overall, our study showed promise in deterring eagles away from reindeer but needs replication in space and in time. Such work is crucial to form a baseline for decision making in natural resource management issues to map and quantify conflicts at different scales.

**Author contributions**

NS, AM, conceived the ideas and designed methodology. NS, AM, CS and PS collected the data. AM and ALP analysed the data. AM and NS led the writing of the manuscript. All authors contributed critically to the drafts and gave final approval for publication.

**Conflict of interest**

Authors declare no conflict of interest at hand.

**Funding**

The study was funded by the Swedish Environmental Protection Agency – Naturvårdsverket.

**Acknowledgements**

We thank Mr. Ante Baer from Vilhelmina Norra Reindeer Herding village for the assistance and advice in the field. County board of Västerbotten and Swedish Environmental Protection Agency provided for financing the project. Linus Ryderfors for help in the field. We also appreciate the help from Kungsörn Sverige and Västerbottens Ornitologiska Förening for their advice on eagle territories and monitoring.

Tables and Figures

**Table 1**. Summary of the Generalized linear mixed model predicting the likelihood of observing eagles (both the Golden eagle and White-tailed eagle combined) in three categories of areas - control, prism and air ventilator. Day is used for accounting for repeated observations. Reference is Control area.

| Variable | Estimate | Std. Error | df | t value | P value |
|---|---|---|---|---|---|
| Intercept | 2.62 | 0.65 | 6 | 3.98 | <0.05 |
| Prism | -1.78 | 0.80 | 12 | -2.21 | <0.05 |
| Ventilator | -1.12 | 0.80 | 12 | -1.38 | 0.19 |

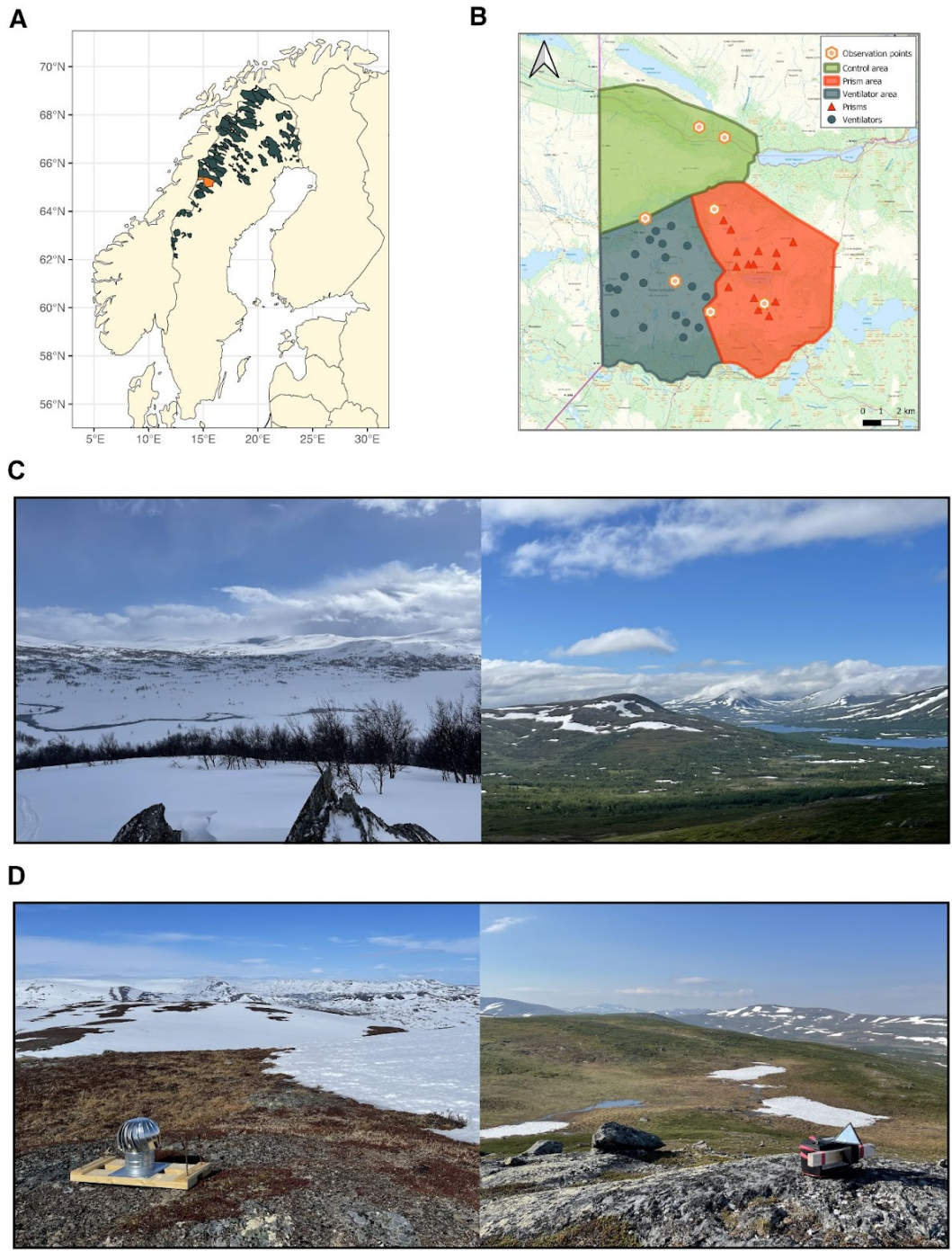

**Figure 1**. Study Area. A: Map of Sweden with reindeer calving areas highlighted (in dark green) and the study area of Vilhelmina Norra (in orange). B: Detailed map of the main calving area within the focus reindeer herding village along with the selected observation points, the delimited areas taken into account for the experiment and the chosen locations for the deterrent devices. C: Examples of landscape in the study area before and after reindeer calving. D: Deterrent devices used in the experiment during the calving period: ventilators (left) and prisms (right).

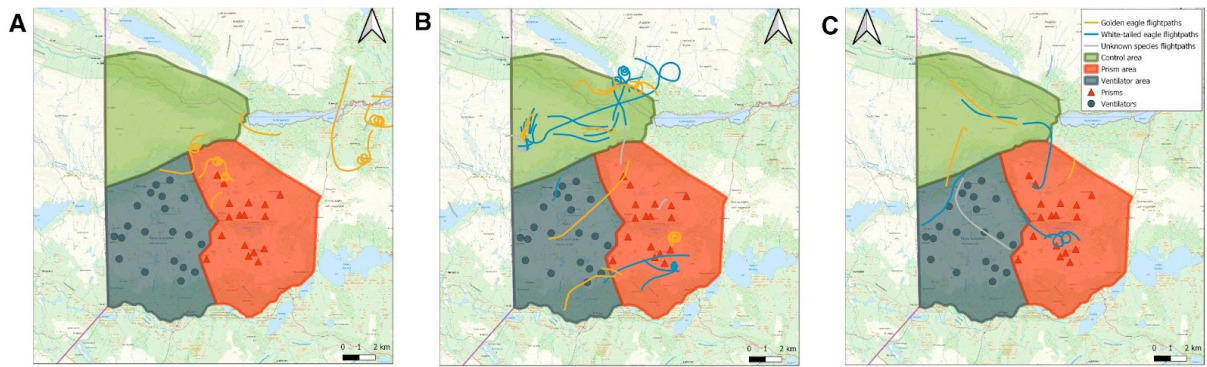

**Figure 2**. Eagle flight patterns at the study area during the three periods: before (A), during (B) and after (C) the calving season. Golden eagles appear in yellow, White-tailed eagles in blue.

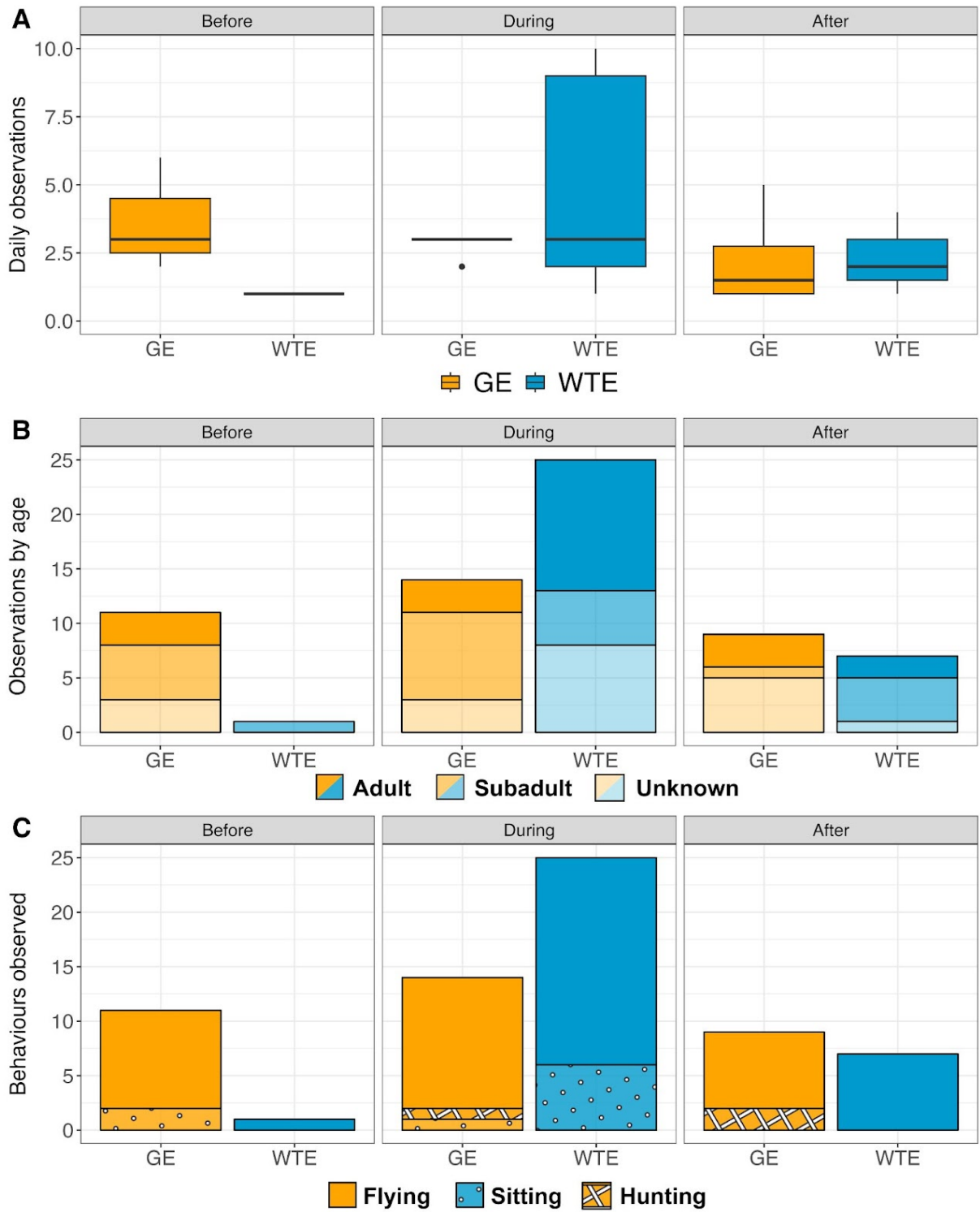

**Figure 3**. Patterns of eagle observations in the study area A: Daily average of Golden eagles - GE (yellow) and White-tailed eagles - WTE (blue) observed during the three observational periods: before, during and after the calving season. B: Eagle observations were split based on age classes: adults (high opacity), subadults (medium opacity) and unknown age (low opacity). C: Eagle observations split by the behaviour: flying (solid pattern), sitting (dot pattern) and hunting (crosshatch pattern).

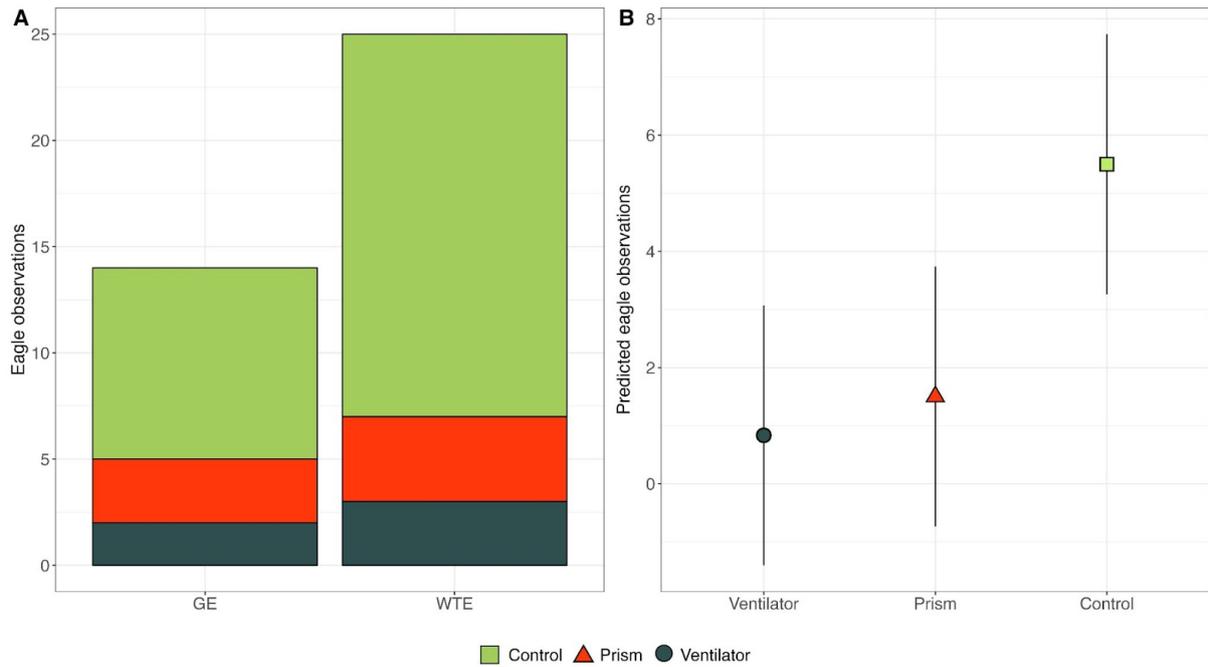

**Figure 4**. Spatial distribution of the eagles during calving. A: Golden eagles - GE (left) and White-tailed eagles - WTE (right) observed during the calving season in the control (light green), prism (red) and ventilator areas (dark green). B: Predicted eagle observations from the GLMM that explained eagle presence in relation to the treatment applied in the area.